\documentclass[article]{IEEEtran}
\usepackage{amsfonts,amsmath,amssymb,epsfig,graphicx,latexsym}
\usepackage{times}
\usepackage[usenames,dvipsnames]{pstricks}
\usepackage{epsfig}
\usepackage{pst-grad} 
\usepackage{pst-plot} 

\newtheorem{theorem}{Theorem}
\newtheorem{lemma}[theorem]{Lemma}

\newtheorem{corollary}[theorem]{Corollary}
\newtheorem{example}[theorem]{Example}

\newcommand\qed{\hfill \rule{1.2mm}{2.8mm}}

\newcommand{\ba}{{\bf a}}
\newcommand{\bb}{{\bf b}}
\newcommand{\bc}{{\bf c}}

\newcommand{\bv}{{\bf v}}
\newcommand{\bw}{{\bf w}}
\newcommand{\x}{{\bf x}}

\newcommand{\mC}{{\mathbb C}}

\newcommand{\mF}{{\mathbb F}}

\newcommand{\be}{\begin{equation}}
\newcommand{\ee}{\end{equation}}

\newcommand{\bea}{\begin{eqnarray*}}
\newcommand{\eea}{\end{eqnarray*}}

\newcommand\wt{\mbox{{\rm wt}\,}}
\newcommand\Tr{\mbox{{\rm Tr}\,}}

\newcommand{\remove}[1]{}

\renewcommand{\le}{\leqslant}\renewcommand{\ge}{\geqslant}

\begin{document}

\title{Fidelity Lower Bounds for Stabilizer and CSS Quantum Codes\thanks{This work was supported by the Intelligence Advanced
Research Projects Activity (IARPA) via Department of Interior
National Business Center contract number D11PC20165.}}

\author{\IEEEauthorblockN{Alexei Ashikhmin, {\em Senior Member, IEEE}}\\
\IEEEauthorblockA{Bell Laboratories, Alcatel-Lucent Inc.\\
Murray Hill, NJ 07974\\
e-mail:~aea@research.bell-labs.com
 }
}

\maketitle

\begin{abstract}
 In this paper we estimate the fidelity of stabilizer and CSS codes. First, we derive a lower bound on the fidelity of a stabilizer code via its quantum enumerator. Next, we find the average quantum enumerators of the ensembles of
finite length stabilizer and CSS codes. We use the average quantum enumerators for obtaining
lower bounds on the average fidelity of these ensembles. We further improve the fidelity bounds by estimating
the quantum enumerators of expurgated ensembles of stabilizer and CSS codes. Finally, we derive fidelity bounds in the asymptotic regime when the code length tends to infinity.

These results tell us which code rate we can afford for achieving a
target fidelity with codes of a given length. The results also show
that in symmetric depolarizing channel a typical stabilizer code has
better performance, in terms of fidelity and code rate, compared
with a typical CSS codes, and that balanced CSS codes significantly
outperform
 other CSS codes. Asymptotic results demonstrate that CSS codes have a fundamental performance loss compared to stabilizer codes.
\end{abstract}

\section{Introduction}
In recent years quantum error correcting codes were subject of intensive studies as they allow protection of
 quantum information from decoherence during quantum computations. The main focus of these studies
 was on
 various constructions of quantum codes, such as block codes, convolutional code, LDPC quantum codes and others,
 and their combinatorial, geometrical and topological properties, such as the minimum distance and
 others (see, for example, numerous publications in {\em IEEE Trans. on
 Information Theory,  Physical Review A, Physical Review Letters, International Journal of Quantum
 Information}).

At the same time it is not so much known about the fidelity that one
can hope to achieve using good quantum codes. A number of lower
bounds on the fidelity $F$ of quantum codes in the asymptotic
regime, as the code length tends to infinity, were derived in
\cite{hamada2001},\cite{hamada2002},\cite{barg}. These bounds were
derived in terms of reliability functions (error exponents).

In this article we consider the problem from a different angle. First, we would like to derive lower bounds on the fidelity $F$ as a
function of the code length. While a reliability function tells us what kind of fidelity we may expect in the asymptotic regime, it does not give a good estimate on the fidelity
for quantum codes of short and moderate code length, like $50$ or $1000$ qubits.

Second, we would like to estimate the fidelity for several ensembles
of quantum codes. In particular, we are interested in the ensembles
of stabilizer codes, linear stabilizer codes,  and CSS codes with
different choices of parameters $k_1$ and $k_2$. CSS codes form an
important subfamily of stabilizer codes. They are attractive for
numerous applications, such as error correction in quantum memory,
quantum fault-tolerant computations, quantum cryptography, and
others. Therefore estimates on their fidelity can be very important
for proper use  of CSS codes in these applications.

Third, we are interested in analysis of performance of CSS codes in the asymptotic regime, as
the code length tends to infinity. In particular, we would like to understand what is the fundamental performance loss of CSS codes compared with the performance of unrestricted  stabilizer codes as the code length tends to infinity.

The paper is organized as follows. In Section
\ref{sec:prelim-depol-chan} we remind the main definitions of
quantum depolarizing channel. In Section \ref{sec:Qenum} we remind
the definitions of classical and quantum enumerators and their main
properties, which will be used later in the paper. In Section
\ref{sec:fidelity_via_enum} we derive a lower bound on the fidelity
$F$ of quantum stabilizer code $Q$ as a function of its quantum
enumerators. Further we derive two lower bounds on the average
fidelity of an ensemble of quantum codes and apply these bounds to
the ensemble of stabilizer codes of given length and code rate.
In Section \ref{sec:F_CSS_codes} we find the average quantum
enumerators of the ensemble of CSS codes and apply them for
obtaining lower bounds on the average fidelity of this ensemble. In
Section \ref{sec:ErrExpon} we investigate the behavior of $F$ for
stabilizer and CSS codes in the asymptotic regime as the code length
tends to infinity.

\section{Preliminaries}\label{sec:prelim}

\subsection{Quantum Depolarizing Channel}\label{sec:prelim-depol-chan}
An $((n,K))$ binary {\em quantum code} $Q$ is a $K$-dimensional
linear subspace of the complex space $\mathbb{C}^{2^n}$.

A {quantum channel} is a trace-preserving completely positive linear
map ${\cal M}$. Any such map has an operator sum representation
$$
{\cal M}(\rho)=\sum_{k\in {\cal K}}  M_k\rho M_k^\dag,
$$
for some operators $M_k$. Here $\rho$ is a density operator on
$\mathbb{C}^{2^{n}}$ and ${\cal K}$ is a set of indices.  We will
write ${\cal M}\sim\{M_k\}_{k\in K}$.

Decoding, or state-recovery operator, ${\cal R}$ associated with $Q$
is another trace-preserving completely positive linear map. The {\em
minimum fidelity} of $Q$ is defined by
\begin{equation}\label{eq:Fidelity_def}
F(Q,{\cal M},{\cal R})=\min_{|\psi \rangle\in Q} \langle \psi |{\cal
R}{\cal M}(|\psi\rangle\langle\psi|)|\psi\rangle.
\end{equation}
In what follows it will be more convenient for us to use the bound
(\ref{eq:HamadaBound}) (see Section \ref{sec:fidelity_via_enum})
instead of working with the definition (\ref{eq:Fidelity_def})
itself. More details on trace-preserving completely positive linear
maps and fidelity of quantum codes can found in
\cite{hamada2001},\cite{hamada2002}, and references within.

The quantum symmetric depolarizing channel is defined with the help
of Pauli operators
$$
\sigma_x=\left[\begin{array}{rr}
0 & 1\\
1 & 0
\end{array}\right], \sigma_z=\left[\begin{array}{rr}
1 & 0\\
0 & -1
\end{array}\right], \sigma_y=\left[\begin{array}{rr}
0 & i\\
-i & 0
\end{array}\right].
$$

Denote by $0,1,\omega,\omega^2$ the elements of the Galois field
$\mathbb{F}_4$. Let us associate with a vector
$\x=(x_1,\ldots,x_n)\in \mathbb{F}_4^n$ the linear operator
$$
E_{\x}=e_1\otimes \ldots\otimes e_n,
$$
where
$$
e_j=\left\{\begin{array}{ll}
I_2, & x_j=0,\\
\sigma_x, & x_j=1,\\
\sigma_z, & x_j=\omega,\\
\sigma_y, & x_j=\omega^2.
\end{array}\right.
$$
The operators $E_\x$ are called {\em error operators}.
 The {\em Hamming weight} of $\x$ is defined in the standard way by
$$
\wt(\x)=|\{x_j\not = 0, j=1,\ldots,n\}|.
$$
The {\em quantum symmetric depolarizing channel} is the channel with
the trace-preserving completely positive linear map
$$
{\cal M}\sim \{\left({p\over
3}\right)^{\small{\wt}(\x)}(1-p)^{n-\small{\wt}(\x)} E_\x:\x\in
\mathbb{F}_4^n\}.
$$
The quantity $p, 0\le p\le 1$, is {\em channel error probability}.

 Equivalently the quantum symmetric depolarizing channel can be
 defined as a channel in which the $j$-th qubit is effected by
 $e_j\in \{ I_2,\sigma_x,\sigma_y,\sigma_z\}$ with probabilities
$$
\Pr(e_j=I_2)=1-p,
$$
$$
\Pr(e_j=\sigma_x)=\Pr(e_j=\sigma_z)=\Pr(e_j=\sigma_y)={p\over 3}.
$$
Thus a quantum code state $|\psi\rangle \in Q$ is effected by the
error operator $E_\x$ with probability
$$
\Pr(E_\x)=\left({p\over
3}\right)^{\small{\wt}(\x)}(1-p)^{n-\small{\wt}(\x)}.
$$

\subsection{Quantum Enumerators}\label{sec:Qenum} Important parameter
of any classical linear code is its {\em weight enumerator}, or its
{\em weight distribution}. The weight enumerator of a linear $[n,k]$
code $C$ over $\mathbb{F}_q$ is defined as the set of numbers:
$$
A_j(C)=|\{\bc\in C:\wt(\bc)=j\}|, j=0,\ldots,n.
$$
The {\em Euclidian dual} code of $C$ is defined by
\begin{align*}
C^\perp=&\{\bc=(c_1,\ldots,c_n):\bc\cdot \bc'=0\\
& \mbox{ for all } \bc'=(c_1',\ldots,c_n')\in C\},
\end{align*}
where
$$
\bc\cdot \bc'=c_1c_1'+\ldots+c_nc_n'
$$
is the Euclidian inner product.

We say that a code $C$ over $\mathbb{F}_4$ is {\em additive} if
$\bc+\bc'\in C$ for any $\bc,\bc'\in C$. The conjugate elements of
$\mathbb{F}_4$ are defined by
$$
\overline{0}=0,\overline{1}=1,\overline{\omega}=\omega^2,\overline{\omega^2}=\omega.
$$
The {\em Hermitian dual code} of an additive code $C$ is defined by
$$
C^\perp=\{\bc:\bc * \bc'=0 \mbox{ for all } \bc'\in C\},
$$
where
\begin{equation}\label{eq:*}
\bc* \bc'=\Tr^{\mathbb{F}_4}_{{\mathbb F}_2} (c_1\overline{
c}_1'+\ldots+c_n\overline{c}_n')
\end{equation}
is the trace inner product. Here $\Tr^{\mathbb{F}_4}_{{\mathbb
F}_2}$ is the trace operator from $\mathbb{F}_4$ into
$\mathbb{F}_2$.  For codes over fields $\mathbb{F}_{q},q> 4,$ their
Hermitian dual codes are defined in \cite{AshKnill}.

In what follows we will use the same notation $C^\perp$ for both
Euclidean and Hermitian dual codes. The meaning will be clear from
the context.

If $C^\perp$ is Euclidean or Hermitian dual of $C$ over
$\mathbb{F}_q$ then its weight enumerator is connected to the weight
enumerator of $C$ via the MacWilliams identities:
$$
A_j^\perp={1\over |C|}\sum_{i=0}^n A_i{\cal K}_j(i),~j=0,\ldots,n,
$$
where
\begin{equation}\label{eq:Kraw}
{\cal K}_j(i)=\sum_{t=0}^j (-q)^t(q-1)^{j-t}{n-t\choose j-t}{i\choose t}
\end{equation}
are Krawtchouk polynomials. Often it is more convenient to formulate
the MacWilliams identities in the following polynomial form. Let
$$
A^\perp(x,y)=\sum_{j=0}^n A_j^\perp x^{n-j}y^j \mbox{ and }
A(x,y)=\sum_{j=0}^n A_j x^{n-j}y^j.
$$
Then
$$
A(x,y)={1\over |C^\perp|}A^\perp(x+(q-1)y,x-y).
$$

In \cite{shor} P.~Shor and R.~Laflamme generalized the notion of
weight enumerators for the case of quantum codes as follows. A
quantum code $Q$ is a linear subspace of $\mathbb{C}^{2^n}$ and
therefore there exists the orthogonal projector $P$ on $Q$. The code
$Q$ has
 two {\em quantum enumerators} $B_j$ and $B_j^\perp$ defined by
\begin{align*}
 B_j^\perp(Q)&={1\over \dim(Q)^2}\sum_{\x\in \mF^n_4:\mbox{\footnotesize wt}(\x)=j}\Tr(E_\x
 P)^2,~j=0,\ldots,n,\\
B_j(Q)&={1\over \dim(Q)}\sum_{\x\in \mF^n_4:\mbox{\footnotesize
wt}(\x)=j}\Tr(E_\x PE_\x P),~j=0,\ldots,n.
\end{align*}

In this paper we are interested only in quantum binary stabilizer
codes. A binary $[[n,k]]$ quantum {\em stabilizer code} $Q$ is a
$2^k$-dimensional linear subspace of $\mC^{2^n}$ that is defined by
 a classical additive code $C$ of length $n$ and  size $2^{n+k}$ over $\mathbb{F}_4$.
The code $C$ has the property that its Hermitian dual $C^\perp$ is a
subset of $C$, that is $C^\perp\subseteq C$.  See
\cite{gott},\cite{stean},\cite{cald} for the exact definition of
stabilizer codes. If $C$ is a linear code over $\mathbb{F}_4$ then
the corresponding quantum code $Q$ is called {\em linear stabilizer}
code.

Denote by
$$
R(C^\perp)={1\over n} \log_4 |C^\perp| \mbox{ and } R(C)={1\over n}\log_4|C|
$$
the code rates of $C^\perp$ and $C$ respectively. These codes rates are connected to
the code rate $R={k\over n}$ of $Q$ by
\begin{equation}\label{eq:R=(Rq+1)over2}
R(C^\perp)={1-R\over 2} \mbox{ and } R(C)={R+1\over 2}.
\end{equation}

The quantum enumerators of a stabilizer code $Q$ are equal to the
enumerators of $C$ and $C^\perp$, that is
\begin{equation}\label{eq:Bj=Aj}
B_j(Q)=A_j(C),~B_j^\perp(Q)=A_j(C^\perp).
\end{equation}
The quantum enumerators of a stabilizer code have a number of useful properties. In particular,
\begin{enumerate}
\item The enumerators $B_j$ and $B_j^\perp$ are nonnegative integers and
$$
B_j\ge B_j^\perp,B_0^\perp=B_0=1.
$$
\item If $t$ is the smallest  integer such that $B_t>B_t^\perp$ then the minimum distance of $Q$ is $d(Q)=t$.
\item  The sum of $B_j$ defines the size of $Q$:
$$
\dim(Q)=2^k={1\over 2^n} \sum_{j=0}^n B_j=2^n\cdot{1\over
\sum_{j=0}^n B_j^\perp}.
$$
\item
Similar to the classical case quantum enumerators are connected to each other via the MacWilliams identities
\begin{equation}\label{eq:MacWill}
B_r={\dim(Q)\over 2^n}\sum_{j=0}^n B_j^\perp {\cal K}_r(j),
\end{equation}
where ${\cal K}_r(j)$ is the quaternary Krawtchouk polynomial
defined in (\ref{eq:Kraw}). In the polynomial form the quantum
MacWilliams identities have the form
\begin{equation}\label{eq:MacWil_polynomial_form}
B(x,y)={\dim(Q)\over 2^n} B^\perp(x+3y,x-y),
\end{equation}
where
\begin{align*}
B^\perp(x,y)&=\sum_{j=0}^n B_j^\perp x^{n-j}y^j, \mbox{ and}\\
B(x,y)&=\sum_{j=0}^n B_j x^{n-j}y^j.
\end{align*}

\end{enumerate}


\section{Bound on Fidelity via Quantum Enumerators}\label{sec:fidelity_via_enum}

\begin{figure}[htb]
\hspace{-28cm}\includegraphics[scale=1]{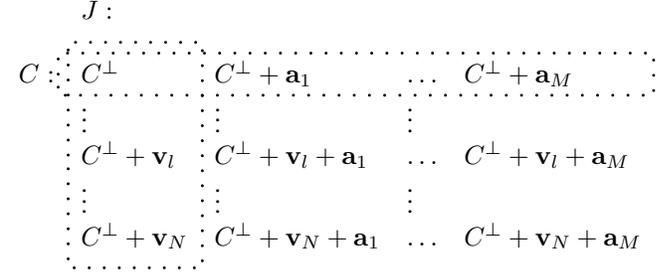} \caption{
Standard Array for Stabilizer Codes. Here $M=2^{2k}-1$ and
$N=2^{n-k}-1$. } \label{fig:stn_array}
\end{figure}

Correctable and uncorrectable errors of a classical linear code in
$q$-ary symmetric channel can be characterized with the help of its
standard array, see for example \cite[Ch.3.3]{blahut}. For a quantum
stabilizer code $Q$ associated with classical code $C$
 one can generalize the standard array as it is shown in Fig.\ref{fig:stn_array}.

Remind that a coset of a linear code $C$ generated by a vector $\bv$ is the set
$$
C+\bv=\{\bc+\bv:\bc\in C\}.
$$
The space $\mathbb{F}_4^n$ can be partitioned into the cosets of
$C$. Each coset of $C$, say $C+\bv_l$,$\bv_l\in\mathbb{F}_4^n$,  can
be further partitioned into cosets of $C^\perp$ for appropriately
chosen coset leaders $\ba_0,\ldots,\ba_{2^{2k}-1}\in C\backslash
C^\perp$. The cosets of $C^\perp$  can be permuted inside the coset
$C+\bv_l$. For example, we can assign $\bv_l'=\bv_l+\ba_j$, and use
$\bv_l'$ instead of $\bv_l$.
   After this permutation the leading (the most left) coset of $C^\perp$ in the $l$-th row of the array will be $C^\perp+\bv_l+\ba_j$.

Any vector ${\bf x}\in \mathbb{F}_4^n$ appears in the standard array one and only one time.
The error operators $E_{{\bf x}}$ that correspond to the vectors ${\bf x}$ from the leading (the most left) cosets
$C^\perp+\bv_l, l=0,\ldots,2^{n-k}-1$
(here we assume $\bv_0=(0,\ldots, 0)$)
form the set $J$ of {\em correctable error operators} (see \cite{gott},\cite{cald} and references within).

\vspace{0.1cm}
\vspace{0.1cm}

In \cite{hamada2001},\cite{hamada2002} M.~Hamada proved, using a
result from \cite{Preskil},
 the following lower bound on {\em minimum average
fidelity} of $Q$:
\begin{equation}\label{eq:HamadaBound}
 F(Q)\ge 1-\sum_{\x \not \in J} \Pr(E_{\x}).
\end{equation}
As we discussed above any coset $C^\perp+\bv_l+\ba_j,j=0,\ldots,2^{2k}-1,$ (here $\ba_0=(0,\ldots,0)$)
can be used as the leading coset in the $l$-th row of the standard array. The optimal choice $j^*$, which maximizes $F(Q)$, is defined by
$$
j^*=\arg\max_{0\le j\le 2^{2k}-1} \sum_{\bc\in C^\perp} \Pr(E_{\bc+\bv_l+\ba_j}).
$$
 The value $j^*$ is not fixed and depends on the channel error
probability $p$. So for different values of $p$ we may have
different $j^*$s. For deriving a lower bound on $F(Q)$ we may choose
any coset $C^\perp+\bv_l+\ba_j$ as the leading one in the $l$-th row
of the standard array. Our goal, of course, is to choose it such
that to get large $F(Q)$. We will use the "classical" approach. Let
$\bw\in C+\bv_l$ be a minimum weight vector in the $l$-th row of the
standard array, that is
$$
\wt(\bw)\le \wt(\bv) \mbox{ for all } \bv\in C+\bv_l.
$$
Then we choose as a leading coset
the one that contains $\bw$, that is we chose $j$ such that
$$
\bw\in C^\perp+\bv_l+\ba_{j}.
$$
Examples show that when $p$ is not too large this is a good and, in
fact, most likely the optimal choice of the leading coset.


In \cite{poltyrev} G.~Poltyrev derived an upper bound on the probability of decoding error of classical linear code via
its weight enumerator $A_j$. The following theorem generalizes this bound for quantum stabilizer codes.
\begin{theorem}\label{thm:poltyrev}
Let $Q$ be a stabilizer code of length $n$ with quantum enumerators $B_j^\perp$ and $B_j$. Then
\begin{align}
&1-F(Q)\nonumber\\
\le& \sum_{m=1}^n \left({p\over 3}\right)^m(1-p)^{n-m}
\min\left\{{n\choose m}3^m, N(m)\right\},\label{eq:poltbound}
\end{align}
where
\begin{equation}\label{eq:N}
N(m)=\sum_{w=1}^{\min(2m,n)} (B_w-B_w^\perp) G(m,w),
\end{equation}
and
\begin{align}
&G(m,w)\nonumber\\
=&\sum_{t=0}^m {w\choose t} 2^t \sum_{h=\lceil {w-t \over 2}\rceil
}^{m-t}  {w-t \choose h} {n-w\choose
m-t-h}3^{m-t-h}.\label{eq:G(m,w)}
\end{align}
\end{theorem}
\IEEEproof
Let us consider an error vector
$$\x=(x_1,\ldots,x_n),\wt(\x)=m.$$
 According to our choice of the leading cosets
in the standard array, the vector $\x$ may belong to $\mathbb{F}_4^n\setminus J$ only if $d(\x,\bc)\le d(\x,{\bf 0})$ for
some $\bc=(c_1,\ldots,c_n)\in C\backslash C^\perp$. Let $\wt(\bc)=w$ and denote
$$
h=|\{ i :x_i\not =0,x_i=c_i\}| \mbox{ and } t=|\{ i: x_i\not =0, x_i\not =c_i\}|.
$$
Then
$$
m-h-t = |\{ i: x_i\not =0, c_i=0\}|.
$$
Using the above notations we get
\begin{align}
d(\x,\bc)&=w-2h-t+m, \label{eq:d}\\
d(\x,{\bf 0})&=m. \label{eq:dx0}
\end{align}
 Using (\ref{eq:d}) and (\ref{eq:dx0}) we get that  the number of vectors $\x$ for which $d(\x,\bc)\le d(\x,{\bf 0})=m$ is equal to $G(m,w)$.

Using union bound we can upper bound the number of error vectors $\x\in \mathbb{F}_4^n\setminus J,\wt(\x)=m$, by $N(m)$.
Taking into account that the total number of error vectors $\x,\wt(\x)=m$, is ${n\choose m}3^m$ we obtain (\ref{eq:poltbound}). \qed



\section{Bounds on the Average Fidelity of an Ensemble of Quantum Codes}\label{sec:F_stab_codes}
In this section we derive bounds on the average fidelity over the
ensemble of stabilizer codes. These bounds can be  considered as
achievability   bounds in the sense that they prove existence of
codes whose fidelity is at least as good as the bounds.

\subsection{Bound on the Average Fidelity}
Denote by ${\cal Q}_{stab}$ {\em the ensemble of all $[[n,k]]$ stabilizer codes}, that is
$$
{\cal Q}_{stab}=\{Q: Q \mbox{ is an $[[n,k]]$ code}\}.
$$
 Let ${\cal Q}$ be an arbitrary sub-ensemble of ${\cal Q}_{stab}$. For instance
${\cal Q}$ can be ${\cal Q}_{stab}$ itself, or it can be the
ensemble of all $(n,k_1,k_2)$ CSS codes (see Section
\ref{sec:F_CSS_codes}).  The average enumerators of ${\cal Q}$ are
defined as
$$
\bar{B}^\perp_j={1\over |{\cal Q}|} \sum_{Q\in {\cal Q}}B^\perp_j(Q),~ \bar{B}_j={1\over |{\cal Q}|} \sum_{Q\in {\cal Q}} B_j(Q).
$$

Similar to the classical case \cite{polyanskiy}[Theorem 5], using the average enumerators $\bar{B}_j^\perp$ and $\bar{B}_j$ in Theorem \ref{thm:poltyrev}, we can obtain an upper bound on the average value of $1-F$ over  the ensemble ${\cal Q}$. We formulate it as a Corollary of Theorem \ref{thm:poltyrev}
\begin{corollary}\label{cor:poltbound_ensembl}
\begin{align}
&1-\bar{F} ={1\over |{\cal Q}|}\sum_{Q\in{\cal Q}} (1-F(Q))\nonumber
\\
\le& \sum_{m=d_{min}/2}^n \left({p\over 3}\right)^m(1-p)^{n-m}
\min\left\{{n\choose m}3^m, \bar{N}(m)\right\}
,\label{eq:poltbound_ensembl}
\end{align}
where
\begin{equation}\label{eq:N(m)}
\bar{N}(m)=\sum_{w=d_{min}}^{\min(2m,n)} (\bar{B}_w-\bar{B}_w^\perp) G(m,w),
\end{equation}
\end{corollary}
\IEEEproof The proof is identical to the classical case. The claim
immediately follows from the observation that the function
$\min\{\cdot,\cdot\}$ is $\bigcap$-convex and therefore Jensen's
inequality can be applied to (\ref{eq:poltbound}). \qed

Corollary \ref{cor:poltbound_ensembl} is an achievability bound in
the sense that it guarantees that in ${\cal Q}$ there exists at
least one code $Q$ with $F(Q)\ge \bar{F}$. Thus studying the average
enumerators of an ensemble of stabilizer codes we get an insight
about the fidelity of the codes from this ensemble.

Denote by ${\cal Q}_{lin.stab}$ the ensemble of all $[[n,k]]$ linear
stabilizer codes, that is
\begin{equation}\label{eq:lin.stab}
{\cal Q}_{lin.stab}=\{Q: Q \mbox{ is a linear $[[n,k]]$ code} \}.
\end{equation}

 The average quantum enumerators of codes
from the ensemble ${\cal Q}_{lin.stab}$ were found in
\cite{ash2000}:
\begin{align}
\bar{B}_0^\perp&=1,\nonumber\\
\bar{B}_j^\perp&={n\choose j}{\alpha\over 2}(3^j+(-3)^j),j\ge
1,\label{eq:stabBperp}\\
\bar{B}_0&=1,\nonumber\\
 \bar{B}_j&={1\over 4^{(n-k)/2}}{n\choose
j}\left((1-\alpha)3^j+{\alpha\over 2}6^j(-2)^{n-j}\right),j\ge
1,\label{eq:stabB}
\end{align}
where
$$
\alpha={4^{(n-k)/2}-1\over {1\over 2}(4^n+(-2)^n)-1}.
$$
  Below we find the average enumerators for
the ensemble ${\cal Q}_{stab}$. Let
$$
{\cal S}_{n,t}=\{C^\perp\subset \mathbb{F}_4^n: |C^\perp|=2^t \mbox{
and } C^\perp\subset C\}
$$
be the ensemble of self-orthogonal codes of length $n$ and size
$2^t$.  We need with the following result.
\begin{lemma}\label{lem:num_of_selforth_codes}
1. The number of codes in ${\cal S}_{n,t}$ is
\begin{equation}\label{eq:|Snt|}
S(n,t)=|{\cal S}_{n,t}|=\prod_{r=0}^{t-1} {2^{2(n-r)}-1\over
2^{r+1}-1}.
\end{equation}

2. Any vector $\bv\in \mathbb{F}_4^n\setminus {\bf 0}$ is contained
in
\begin{equation}\label{eq:|v in Snt|}
L(n,t)=\prod_{r=1}^{t-1} {2^{2(n-r)}-1\over 2^{r}-1}
\end{equation}
codes from ${\cal S}_{n,t}$.
\end{lemma}
\IEEEproof Let $C_r^\perp\in {\cal S}_{n,r}$ and $C_r$ be the dual
of $C^\perp_r$. Any vector $\bw\in \mathbb{F}_4^n$ is
self-orthogonal with respect to the trace inner product
(\ref{eq:*}), i.e. $\bw*\bw=0$. Hence if we take any $\bw\in
C_r\setminus C_r^\perp$ then the code
$C^\perp_{r+1}=C^\perp_r\bigcup (C^\perp_r+\bw)$ is again
self-orthogonal. Using instead of $\bw$ any other vector from the
coset $(C^\perp_r+\bw)$ we obtain the same code $C_{r+1}^\perp$.
Since $|C_r\setminus C_r^\perp|=2^{2n-r}-2^{r}$ we can construct
$(2^{2n-r}-2^{r})/2^r$ different codes $C^\perp_{r+1}$ from a given
$C_r^\perp$. In $C_{r+1}^\perp$ there exist $2^{r+1}-1$ different
subcodes of size $2^r$. Hence
$$
S(n,r+1)={2^{2n-r}-2^r\over 2^r(2^{r+1}-1)} S(n,r).
$$
From this expression (\ref{eq:|Snt|}) follows.

Let $C_1^\perp=\{{\bf 0},\bv\}$. This code is self-orthogonal and
can be used as a starting point for construction of codes
$C_{r+1}^\perp$  in the above procedure. The only difference from
the above procedure is that in $C_{r+1}^\perp$ there exist $2^r-1$
codes $C_r^\perp$ such that $C_1^\perp\subset C_r^\perp$ (see Lemma
\ref{lem:C1s_with_vector_a_in_given_C2}). Taking this into account
we obtain the expression (\ref{eq:|v in Snt|}). \qed

Now we can find the average enumerators of stabilizer codes.
\begin{theorem} For the ensemble ${\cal Q}_{stab}$ we have
\begin{equation}
 \bar{B}_0^\perp=1,
 \bar{B}_j^\perp ={2^{n-k}-1\over 4^n-1}{n\choose
 j}3^j,~j\ge 1,\label{Bjperp_general.stab}
 \end{equation}
 \begin{equation}
 \bar{B}_0=1,
 \bar{B}_j={2^{n+k}-1\over 4^n-1}{n\choose j}3^j, j\ge 1.
 \label{eq:Bj_general.stab}
 \end{equation}
 \end{theorem}
\IEEEproof Because of  (\ref{eq:Bj=Aj}), the enumerator
$\bar{B}^\perp_j$ coincides with the average enumerator of classical
codes from the ensemble ${\cal S}_{n,n-k}$, which is easy to find.
Indeed,  the number of vectors $\bw\in \mathbb{F}_4^n$ of weight $j$
is ${n\choose j}3^j$. From Lemma \ref{lem:num_of_selforth_codes} we
have
$$
{L(n,n-k)\over S(n,n-k)}={2^{n-k}-1\over 4^n-1},
$$
and the expression (\ref{Bjperp_general.stab}) follows.

Let us denote $\gamma={2^{n-k}-1\over 4^n-1}$. It is easy to see
that
$$
\bar{B}^\perp(x,y)=\sum_{j=0}^n \bar{B}^\perp_j y^j
x^{n-j}=(1-\gamma)x^n+\gamma (x+3y)^n.
$$
Using the quantum MacWilliams identities
(\ref{eq:MacWil_polynomial_form}), we obtain
\begin{align*}
\bar{B}(x,y)&=\sum_{j=0}^n \bar{B}_j y^j
x^{n-j}\\
&=2^{k-n}\left((1-\gamma)(x+3y)^n+\gamma (4x)^n\right).
\end{align*}
From this the expression (\ref{eq:Bj_general.stab}) follows. \qed

Using (\ref{Bjperp_general.stab}) and (\ref{eq:Bj_general.stab}) in
(\ref{eq:poltbound_ensembl}) we obtain a bound on $1-\bar{F}$ for
the ensemble ${\cal Q}_{stab}$. This bound for $n=50$ and $k=22$ is
presented in Fig.\ref{fig:Perr_50_14}.

\subsection{Bound on the Average Fidelity of Expurgated Ensemble}
Below we show that for small values of the channel error $p$ one can significantly improve the bound (\ref{eq:poltbound_ensembl}). In order of doing this we first note that after simple algebraic manipulations the bound (\ref{eq:poltbound}) can be transformed to the following form.

Let $m_0$ be the smallest integer such that
\begin{equation}\label{eq:Nm1}
N(m_0)\ge {n\choose m_0}3^{m_0}.
\end{equation}
and let
\begin{align}
T_w&=(B_w-B_w^\perp) \sum_{t=0}^{m_0-1} {w\choose t} 2^t
\sum_{h=\lceil
{w-t \over 2}\rceil }^{m_0-t-1} {w-t \choose h}\nonumber \\
& \times \sum_{\kappa=0}^{m_0-1-t-h}  {n-w\choose \kappa}3^{\kappa}
(p/3)^{\kappa+t+h}(1-p)^{n-\kappa-t-h} \label{eq:T}
\end{align}
Then
\begin{equation}\label{eq:poltyrev1}
1-F\le \sum_{w=1}^{2(m_0-1)} T_w + \sum_{l=m_0}^n {n\choose l}
(p/3)^l (1-p)^{n-l}.
\end{equation}

Define $\bar{T}_w$ as in (\ref{eq:T}), but replacing ${B}_j$ and ${B}_j$ with $\bar{B}_j$ and $\bar{B}_j$ respectively.
  The right hand side of (\ref{eq:poltyrev1}) is equal to the right hand side of (\ref{eq:poltbound}). Hence if we  use $\bar{N}(m)$ (defined in (\ref{eq:N(m)})) in (\ref{eq:Nm1}) and $\bar{T}_w$  in (\ref{eq:poltyrev1}) we obtain
\begin{equation}\label{eq:poltyrev1_ensemble}
1-\bar{F}\le \sum_{w=d_{min}}^{2(m_0-1)} \bar{T}_w + \sum_{l=m_0}^n
{n\choose l} (p/3)^l (1-p)^{n-l}.
\end{equation}

\begin{figure}[htb]
\centering
\includegraphics[scale=0.35]{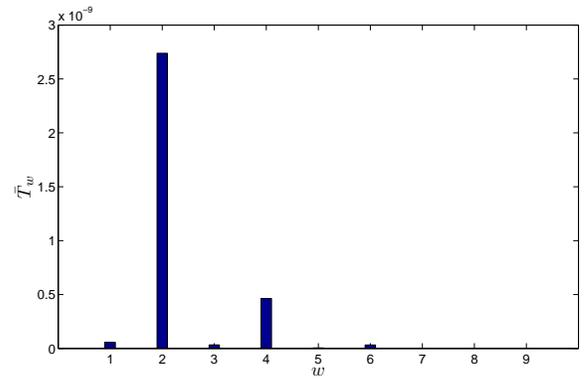}
\caption{$\bar{T}_w$ for the ensemble of all $[[50,22]]$ stabilizer codes}
\label{fig:Tw}
\end{figure}

Let us now examine the individual contributions of the  terms of (\ref{eq:poltyrev1_ensemble}).
 For  instance, in the case of the ensemble of all $[[50,22]]$ stabilizer codes and $p=10^{-4}$ the values of  $\bar{T}_w$
 are shown in Fig.\ref{fig:Tw}.
One can see that the main contributions are made by $\bar{T}_2$ and
$\bar{T}_4$.

In \cite{feng2004} the following lower bound on the minimum distance of stabilizer (pure) codes was proven.
\begin{theorem}\label{thm:GV_stab} \cite{feng2004} For $n>k\ge 2,d\ge 2$ and $n=k\mod 2$ there exists a pure stabilizer code $[[n,k,d_{GV}(n,k)]]$ provided
$$
{2^{n-k+2}-1\over 3} > \sum_{i=1}^{d_{GV}(n,k)-1} 3^{i-1}{n\choose
i}.
$$
\end{theorem}
(For $n\not = k \mod 2$ a similar bound can be derived.)
  According
to this bound there exist $[[50,22]]$ codes with $d\ge 5$. For such
codes, using the property 2 of $B_j$ and $B_j^\perp$ from Section
\ref{sec:Qenum}, we get
$$
\bar{T}_1=\bar{T}_2=\bar{T}_3=\bar{T}_4=0.
$$
So, one can hope that codes with $d\ge 5$ will have significantly
smaller $1-\bar{F}$ than an average $[[50,22]]$ code.

Let ${\cal Q}$ be a sub-ensemble of ${\cal Q}_{stab}$ and let $\bar{B}_j^\perp$ and $\bar{B}_j$ be
the average enumerators of ${\cal Q}$. Assume that there is at least one $j$ such that $\bar{B}_j-\bar{B}_j^\perp <1$ and
consider a set
$$
I \subseteq \{j: \bar{B}_j-\bar{B}_j^\perp <1\}.
$$
Define the expurgated ensemble of ${\cal Q}_{ex}$ as
$$
{\cal Q}_{ex} =\{Q\in {\cal Q}: B_j(Q)-B_j^\perp(Q)=0 \mbox{ for all }j\in I\}.
$$
For upper-bounding the average value of $1-F$ over ${\cal Q}_{ex}$,
we have to estimate the average quantum enumerators of ${\cal
Q}_{ex}$. The following theorem gives an upper bound on the average
quantum enumerators of ${\cal Q}_{ex}$ in terms of the average
enumerators of the original ensemble ${\cal Q}$.
\begin{theorem}\label{thm:enumerators_of_Q_I}
Let $\hat{B}_j^\perp$ and $\hat{B}_j$ be the average quantum enumerators of ${\cal Q}_{ex}$ and let
$$
\beta=1-\sum_{j\in I} (\bar{B}_j-\bar{B}_j^\perp).
$$
If $\beta>0$ then ${\cal Q}_{ex}$ is not empty and
$$
\hat{B}_j-\hat{B}_j^\perp\le {1\over \beta}
(\bar{B}_j-\bar{B}_j^\perp),~j\not\in I.
$$
\end{theorem}
\IEEEproof Let $Q$ be a randomly chosen code from ${\cal Q}$ with
respect to the uniform distribution. Then from Markov's inequality
it follows that for a positive $\alpha_j$ we have
$$
\Pr\left((B_j(Q)-B_j^\perp(Q))\ge \alpha_j (\bar{B}_j-\bar{B}_j^\perp)\right)\le {1\over \alpha_j}.
$$
For $I=\{j_1,\ldots, j_m\}\subseteq \{j: \bar{B}_j-\bar{B}_j^\perp
<1\}$, using the union bound, we obtain
\begin{align}
&\Pr\left(B_{j_r}(Q)-B_{j_r}^\perp(Q)\le \alpha_{j_r}
(\bar{B}_{j_r}-\bar{B}_{j_r}^\perp)\mbox{ for all }j_r\right)\nonumber\\
 \ge&
1-{1\over \alpha_{j_1}}-\ldots -{1\over \alpha_{j_m}}.
\label{eq:beta}
\end{align}
Let
$$
\alpha_{j_r}={1-\epsilon \over \bar{B}_{j_r}-\bar{B}^\perp_{j_r}},
\epsilon>0.
$$
If $\beta>0$ then we can choose $\epsilon>0$ such that the right
hand side of (\ref{eq:beta}) is positive. Hence there exists $Q$
such that
$$
B_{j_r}(Q)-B_{j_r}^\perp(Q)\le \alpha_{j_r}
(\bar{B}_j-\bar{B}_j^\perp),\mbox{ for all }j_r\in I.
$$
Since $\alpha_{j_r}(\bar{B}_{j_r}-\bar{B}^\perp_{j_r})<1$ and
$B_j(Q)-B_j^\perp(Q)$ are integers (according to the first property
of quantum enumerators in Section \ref{sec:Qenum}), we have
$$
B_{j_r}(Q)-B_{j_r}^\perp(Q)=0,\mbox{ for all }j_r\in I.
$$
Summarizing this, we conclude that there exists a nonempty ensemble
$$
{\cal Q}_{ex}\subseteq {\cal Q}
$$
of stabilizer codes and for $Q\in {\cal Q}_{ex}$ we have
$B_j(Q)-B_j^\perp(Q)=0,j\in I$.

From (\ref{eq:beta})
 we have $|{\cal Q}_{ex}|\ge \lceil \beta |{\cal Q}|\rceil$. Hence we can get the following upper
 bound on $\hat{B}_j-\hat{B}_j^\perp$:
\begin{align*}
&\bar{B}_j-\bar{B}_j^\perp\nonumber\\
=&{1\over |{\cal Q}|} \left( \sum_{Q\in {\cal Q}_{ex}}
(B_j(Q)-B_j^\perp(Q))
 +\sum_{Q\not \in {\cal Q}_{ex}} (B_j(Q)-B_j^\perp(Q))\right)\\
 \ge&  {1\over |{\cal Q}|} \sum_{Q\in {\cal Q}_{ex}} (B_j(Q)-B_j^\perp(Q))\\
 =&{|{\cal Q}_{ex}|\over |{\cal Q}|} (\hat{B}_j-\hat{B}_j^\perp)
 \ge  \beta (\hat{B}_j-\hat{B}_j^\perp).
 \end{align*}
\qed

\begin{example}
Let ${\cal Q}$ be the ensemble of all $[[50,22]]$ stabilizer codes. Then we have
\begin{eqnarray*}
 \bar{B}_1-\bar{B}_1^\perp\approx 5.6\cdot 10^{-5},&&\bar{B}_2-\bar{B}_2^\perp\approx 4\cdot 10^{-5},\\
 \bar{B}_3-\bar{B}_3^\perp\approx 2\cdot 10^{-3},&&\bar{B}_4-\bar{B}_4^\perp\approx 6.9\cdot 10^{-2}.\\
\end{eqnarray*}
 Let us form ${\cal Q}_{ex}$ with $I=\{1,2,3,4\}$. In this case we have $\beta> 0.9285$.
 Thus for the expurgated ensemble ${\cal Q}_{ex}$ we have
\begin{eqnarray*}
 \hat{B}_1-\hat{B}_1^\perp=0,&&\hat{B}_2-\hat{B}_2^\perp=0,\\
 \hat{B}_3-\hat{B}_3^\perp=0,&&\hat{B}_4-\hat{B}_4^\perp=0.\\
\end{eqnarray*}
and for $j>4$
$$
(\hat{B}_j-\hat{B}_j^\perp) \le 1.077\cdot (\bar{B}_j-\bar{B}_j^\perp).
$$
Upper bounds on $1-\bar{F}$ for ${\cal Q}$ and ${\cal Q}_{ex}$ are shown in
fig.\ref{fig:Perr_50_14}.

\begin{figure}[htb]
\centering
\hspace{-.1cm} \includegraphics[scale=0.38]{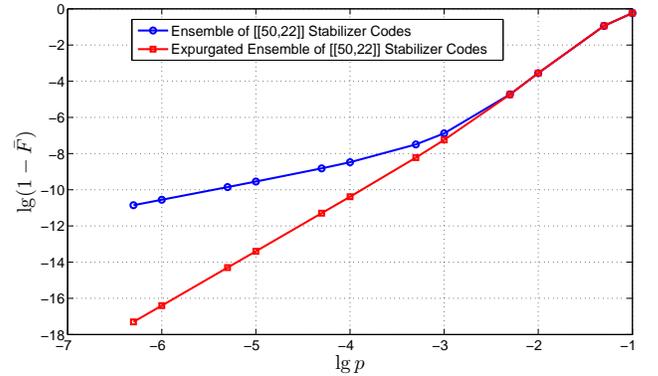}
\caption{Lower bounds on $1-\bar{F}$ for ${\cal Q}_{stab}$ and ${\cal Q}_{stab,ex}$}
\label{fig:Perr_50_14}
\end{figure}

One can see that when $p<10^{-3}$ the bound for the expurgated ensemble ${\cal Q}_{ex}$ is getting significantly
 better than the bound for ${\cal Q}_{stab}$.
\end{example}

\noindent{\bf Remark} Typically the average enumerators
(\ref{Bjperp_general.stab}) and (\ref{eq:Bj_general.stab}) of
stabilizer codes are very close to the average enumerators
(\ref{eq:stabBperp}) and (\ref{eq:stabB}) of linear stabilizer
codes. This results in that the fidelity bounds for these ensembles
are also basically identical. All results presented for stabilizer
codes in Fig.\ref{fig:Perr_50_14}, \ref{fig:1_F}, and
\ref{fig:1-Fgrowing_n}, and are also valid for linear stabilizer
codes with the same parameters.

\section{Bounds on the Fidelity of CSS Codes}\label{sec:F_CSS_codes}

CSS codes \cite{css1},\cite{css2} form an important subclass of stabilizer codes. CSS codes are good candidates for some
practical applications, such as quantum cryptographical protocols and  fault-tolerant quantum computations.
Hence it looks natural to try to estimate
their performance and compare it with the performance of unrestricted stabilizer codes.

CSS codes form a subfamily of stabilizer codes. It is more
convenient to define them with the help of binary classical codes,
rather than classical codes over ${\mathbb F}_4$.

Remind that any vector $\bv\in {\mathbb F}_4^n$ can be written in
the form $\bv=\ba+\omega\bb$, where
$\ba=(a_1,\ldots,a_n),\bb=(b_1,\ldots,b_n)\in {\mathbb F}_2^n$. Note
that
\begin{equation}\label{eq:bin_wt}
\wt(\bv)=|\{j:(a_j,b_j)\not=(0,0)\}.
\end{equation}
(See \cite{css1},\cite{cald} for details).

An $(n,k_1,k_2)$ CSS code is defined by two classical binary codes
$C_1$ and $C_2$ with the property that $C_1\subseteq C_2$. The code
$C_1$ is an $[n,k_1]$ code and code $C_2$ is an $[n,k_2]$ code. By
$C_1^\perp$ and $C_2^\perp$ we denote the Euclidean dual codes of
$C_1$ and $C_2$ respectively. The CSS is a stabilizer code. The
generator matrix of its associated code $C^\perp$ (in the binary
form) is defined by
\begin{equation}\label{eq:G_of_CSS}
G_{C^\perp}=\left[\begin{array}{ll}
G_{C_1} & 0\\
0 & G_{C_2^\perp}
\end{array}\right],
\end{equation}
where $G_{C_1}$ and $G_{C_2^\perp}$ are generator matrices of $C_1$ and $C_2^\perp$ respectively.

We will say that a CSS code is {\em balanced} if $k_1=n-k_2$.

Define the ensemble of CSS codes as
$$
{\cal Q}_{CSS}=\{Q: Q \mbox{ is an $(n,k_1,k_2)$ CSS code}\}.
$$

We need the following lemmas for deriving the average enumerators of
${\cal Q}_{CSS}$. The proofs of the Lemmas are in the Appendix.

Remind that the binary {\em Gaussian binomial coefficients} are defined by
$$
\left[\begin{array}{cc}
n\\
0
\end{array}\right]=1,
$$
$$
\left[\begin{array}{cc}
n\\
k
\end{array}\right]={(2^n-1)(2^{n-1}-1)\ldots (2^{n-k+1}-1)\over (2^k-1)(2^{k-1}-1)\ldots
(2-1)},
$$
and that
$$
\left[\begin{array}{cc}
n\\
n-k
\end{array}\right]=\left[\begin{array}{cc}
n\\
k
\end{array}\right].
$$

\begin{lemma}\label{lem:1}
Let $\bv,\bw\in \mathbb{F}_2^n\setminus{\bf 0}$ and $\bv\cdot
\bw=0$. If $\bv \in C_1$ and $\bw\in C_2^\perp$ then code $C_2$ can
be chosen in
$$
\left[\begin{array}{cc}
n-2\\
k_2-1
\end{array}\right]
$$
ways.
\end{lemma}

\begin{lemma}\label{lem:C2s that contain given C1}
The number of $[n,k_2]$ codes that contain a given $[n,k_1]$ code is equal to
$$
\left[\begin{array}{cc}
n-k_1\\
k_2-k_1
\end{array}\right].
$$
\end{lemma}

\begin{lemma}\label{lem:C1s codes in a given C2}
The number of $[n,k_1]$ codes in a given $[n,k_2]$ code is equal to
$$
\left[\begin{array}{cc}
k_2\\
k_1
\end{array}\right].
$$
\end{lemma}

\begin{lemma}\label{lem:C1s_with_vector_a_in_given_C2}
Let $\bv\in {\mathbb F}_2^n\setminus {\bf 0}$. The number of
$[n,k_1]$ codes $C_1$ in a given $[n,k_2]$ code such that $\bv\in
C_1$ is equal to
$$
\left[\begin{array}{cc}
k_2-1\\
k_1-1
\end{array}\right].
$$
\end{lemma}

\begin{lemma}\label{lem:w in C2perp}
Let $C_2$ be an $[n,k_2]$ code and let $\bw\in C_2^\perp$. Then $C_2$ can be constructed in
$$
\left[\begin{array}{cc}
n-1\\
k_2
\end{array}\right].
$$
ways.
\end{lemma}

Using the above Lemmas we will prove the following theorems that define the average distance distribution of $(n,k_1,k_2)$ CSS codes.
\begin{theorem}\label{thm:(1010|0000) vectors}
Let $\bv=(\ba,{\bf 0})\in \mF_2^{2n}$, where ${\bf
0}=(0,\ldots,0)\in \mathbb{F}_2^n$, and $\ba\in
\mathbb{F}_2^n\setminus {\bf 0}$. Then $\bv$ is contained in
$$
\left[\begin{array}{cc}
n-1\\
k_1-1
\end{array}\right]
\left[\begin{array}{cc}
n-k_1\\
k_2-k_1
\end{array}\right]
$$
codes with generators matrices of the form (\ref{eq:G_of_CSS}). The
number of such vectors is $2^n-1$.
\end{theorem}
\IEEEproof The number of $C_1$ codes that contain the vector $\ba$
is
$$
\left[\begin{array}{cc}
n-1\\
k_1-1
\end{array}\right].
$$
The number of $C_2$ codes that contain a given $C_1$ is defined in Lemma \ref{lem:C2s that contain given C1}.
This finishes the proof. \qed

\begin{theorem}\label{thm:(0000|1010)}
Let $\bv=({\bf 0},\ba)\in \mF_2^{2n}$, where ${\bf
0}=(0,\ldots,0)\in \mathbb{F}_2^n$, and $\ba\in
\mathbb{F}_2^n\setminus {\bf 0}$. Then $\bv$ is contained in
$$
\left[\begin{array}{cc}
n-1\\
k_2
\end{array}\right]
\left[\begin{array}{cc}
k_2\\
k_1
\end{array}\right]
$$
codes with generators matrices of the form (\ref{eq:G_of_CSS}). The
number of such vectors is $2^n-1$.
\end{theorem}
\IEEEproof The number of codes $C_2$ such that $\ba\in C_2^\perp$ is
defined by Lemma \ref{lem:w in C2perp}. The number of $C_1$ codes in
a given $C_2$ is determined by Lemma \ref{lem:C1s codes in a given
C2}. \qed

\begin{theorem}\label{thm:vectors with even number of y}
Let $\ba,\bb \in \mathbb{F}_2^n\setminus {\bf 0}$ and let
$\bv=(\ba,\bb)\in \mF_2^{2n}$ be a vector such that
$|\{a_j=b_j=1\}|$ is a positive even number. Then $\bv$ is contained
in
$$
\left[\begin{array}{cc}
n-2\\
k_2-1
\end{array}\right]
\left[\begin{array}{cc}
k_2-1\\
k_1-1
\end{array}\right]
$$
codes with generators matrices of the form (\ref{eq:G_of_CSS}). The
number of such vectors is
$$
\sum_{i>0,~i \mbox{ {\small is even}}}^n {n\choose i}3^{n-i}.
$$
\end{theorem}
\IEEEproof Since $|\{a_j=b_j=1\}|$ is even we have $\ba\cdot\bb=0$.
Hence we can construct
CSS codes such that $\ba\in C_1$ and $\bb\in C_2^\perp$. The number
of $C_2^\perp$ codes of this type, and therefore the number of $C_2$
codes, is defined by Lemma \ref{lem:1}. The number of $C_1$ codes
with $\ba\in C_1$ that are contained in a given $C_2$ is defined by
Lemma \ref{lem:C1s_with_vector_a_in_given_C2}. \qed

\begin{theorem}\label{thm:vectors with odd number of y}
Let $\ba,\bb \in \mathbb{F}_2^n\setminus {\bf 0}$ and let
$\bv=(\ba,\bb)\in \mF_2^{2n}$ be a vector such that
$|\{a_j=b_j=1\}|$ is an odd number. Then $\bv$ does not belong to
any code with generator matrix of the form (\ref{eq:G_of_CSS}).
\end{theorem}
\IEEEproof If $\bv=(\ba,\bb)$, with nonzero $\ba$ and $\bb$, then
$\ba\in C_1$ and $\bb\in C_2^\perp\subseteq C_1^\perp$. At the same
time $|\{a_j=b_j=1\}|$ is odd and therefore $\ba\cdot \bb =1$. A
contradiction. \qed
\begin{theorem}\label{thm:xz}
Let $\ba,\bb\in \mathbb{F}_2^n\setminus {\bf 0}$ and there is no $j$ with $a_j=1,b_j=1$. Then the vector
$(\ba,\bb)$ is contained in
 $$
\left[\begin{array}{cc}
n-2\\
k_2-1
\end{array}\right]
\left[\begin{array}{cc}
k_2-1\\
k_1-1
\end{array}\right]
$$
codes with generators matrices of the form (\ref{eq:G_of_CSS}). The
number of such vectors is
$$3^n-2(2^n-1).$$
\end{theorem}
\IEEEproof Since there is no $j$ such that $a_j=1$ and $b_j=1$ we
have $\ba\cdot\bb=0$.
Hence we can construct CSS codes such that $\ba\in C_1$ and $\bb\in
C_2^\perp$. The number of such CSS codes is defined by Lemmas
\ref{lem:1} and \ref{lem:C1s_with_vector_a_in_given_C2}. \qed

\begin{theorem}\label{thm:wt_distCSS}
The average quantum enumerators of the ensemble ${\cal Q}_{CSS}$  is
$$
\bar{B}_0^\perp=1,
$$
\begin{equation}\label{eq:Bperp}
\bar{B}_j^\perp=
{1\over c_1}\left[{n\choose j}(c_2-{3\over 2}c_3)+{c_3\over 2}{n\choose j}3^j\right],  j\ge 1.
\end{equation}
and
$$
\bar{B}_0=1,
$$
\begin{align}
\bar{B}_j &= {1\over 2^{k_1+n-k_2}c_1} \nonumber\\
& \cdot \left({n\choose j}[2^n(c_2-{3\over
2}c_3)+3^j(c_1-c_2+c_3)]\right),j\ge 1,\label{eq:B}
\end{align}
where \begin{align*}
 c_1&=\left[\begin{array}{c}
n\\
k_2\end{array}\right]
\left[\begin{array}{c}
k_2\\
k_1
\end{array}\right],\\
c_2&=\left[\begin{array}{c}
n-1\\
k_1-1\end{array}\right]
\left[\begin{array}{c}
n-k_1\\
k_2-k_1
\end{array}\right]
+
\left[\begin{array}{c}
n-1\\
k_2
\end{array}\right]
\left[\begin{array}{c}
k_2\\
k_1
\end{array}\right],\\
c_3&=\left[\begin{array}{c}
n-2\\
k_2-1
\end{array}\right]
\left[\begin{array}{c}
k_2-1\\
k_1-1
\end{array}\right].
\end{align*}
\end{theorem}
\IEEEproof
 Remind that the weight of a vector $\ba+\omega\bb$ can be
found according to (\ref{eq:bin_wt}).

 It is easy to see that the number
of vectors from Theorem \ref{thm:(1010|0000) vectors} of weight $j$
is ${n \choose j}$. The same is true for vectors from Theorem
\ref{thm:(0000|1010)}. In the case of Theorem \ref{thm:vectors with
even number of y} we have that the number of vectors of weight $j$
is
$$
\sum_{i>0,~i~ \mbox{\small is even}}^j {n\choose i}{n-i\choose
j-i}2^{j-i}
$$
and for Theorem \ref{thm:xz} this number is
$$
{n\choose j}(2^j-2).
$$
Combining these results we obtain
\begin{align*}
\bar{B}_j^\perp&={1\over c_1} \left[{n\choose j}c_2\right.\\
&+ \left.\left(\sum_{i>0,~i \mbox{ {\small is even}}}^j {n\choose
i}{n-i\choose j-i}2^{j-i}+{n\choose j}(2^j-2)\right)c_3\right].
\end{align*}
Using the identities
$$
{n\choose i}{n-i\choose j-i}={n\choose j}{j\choose i}
$$
and
$$
\sum_{i=0,~i\mbox{ \small is even}}^j {j\choose i}2^{-i}= {1\over
2}\left[\left(1+{1\over 2}\right)^j + \left(1-{1\over
2}\right)^j\right],
$$
after simple manipulations, we obtain (\ref{eq:Bperp}).

The generating function of  the $q$-ary Krawtchouk polynomials (see \cite[Ch.5.7]{macwilliams})
is
$$
(1+(q-1) z)^{n-j}(1-z)^j =\sum_{r=0}^n {\cal K}_r(j)z^r.
$$
Hence, in the case $q=4$, the sum
$$
\sum_{j=0}^n {n\choose j} {\cal K}_r(j)
$$
is equal to the coefficient of $z^r$ of the polynomial
$$
\sum_{j=0}^n {n\choose j}(1+3z)^{n-j}(1-z)^j=2^n(1+z)^n.
$$
Thus for $q=4$ we have
$$
\sum_{j=0}^n {n\choose j} {\cal K}_r(j)=2^n{n\choose r}.
$$
In a similar way  it is easy to show that
$$
\sum_{j=0}^n {n\choose j}3^j {\cal K}_r(j)=4^n\delta_{0,r}.
$$
Using these expressions together with the MacWilliams identities (\ref{eq:MacWill}) we obtain (\ref{eq:B}). \qed

For a set
$$
I\subseteq \{ \bar{B}_j(Q)-\bar{B}_j^\perp(Q)<1\}
$$
we define the expurgated ensemble
$$
{\cal Q}_{CSS,ex}=\{Q\in {\cal Q}_{CSS}:B_j(Q)-B_j^\perp(Q)=0 \mbox{ for all }j\in I\}.
$$
The average quantum enumerators of ${\cal Q}_{CSS,ex}$ can be upper bounded with the help of Theorems \ref{thm:enumerators_of_Q_I} and \ref{thm:wt_distCSS}.

Using Theorem \ref{thm:wt_distCSS} and the union bound we obtain  a
Gilbert-Varshamov type bound for finite length CSS codes.
\begin{theorem}\label{thm:GV_CSS}
Let $d_{GV}(n,k_1,k_2)$ be the largest integer such that
$$
1-\sum_{j=1}^{d_{GV}(n,k_1,k_2)-1}
\left(\bar{B}_{j}-\bar{B}^\perp_{j}\right)>0.
$$
Then there is exists an $(n,k_1,k_2)$ CSS code with minimum distance
$d_{GV}(n,k_1,k_2)$.
\end{theorem}

\begin{figure}[htb]
\centering
\hspace{-0.5cm}\includegraphics[scale=0.38]{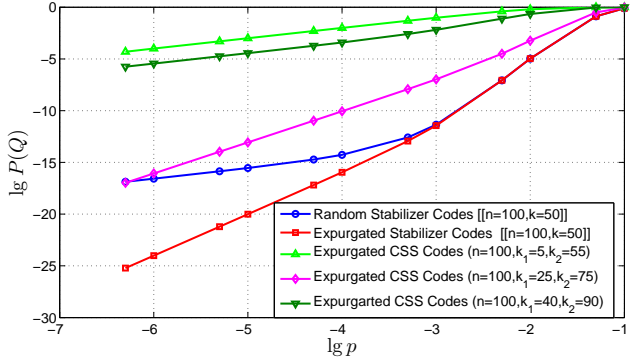}
\caption{Lower bounds on $1-\bar{F}$ for the ensemble of stabilizer
codes and different ensembles of CSS codes; all codes have the same
length $n=100$ and code rate $R=1/2$.} \label{fig:1_F}
\end{figure}

In Fig.\ref{fig:1_F} the quantity $1-\bar{F}$ for ensembles of
${\cal Q}_{stab}$, ${\cal Q}_{stab,ex}$, and ${\cal Q}_{CSS,ex}$ is
presented. All considered codes have the same length $n=100$ and the
same code rate $R=1/2$. One can see that codes from ${\cal
Q}_{stab,ex}$ start to significantly outperform codes from ${\cal
Q}_{stab}$ only at $p<10^{-3}$. For larger values of $p$ the
expurgation does not play any role. This result can also be
interpreted as that for $p>10^{-3}$ the minimum distance of a
typical $[[50,22]]$ code does effect its fidelity.  Another
observation is that stabilizer codes are significantly better than
CSS codes. Among CSS codes the balanced code ($k_1=n-k_2$) happened
to have significantly better performance than codes with
$k_1<<n-k_2$ or $n-k_2<<k_1$.

\begin{figure}[htb]
\centering
\includegraphics[scale=0.38]{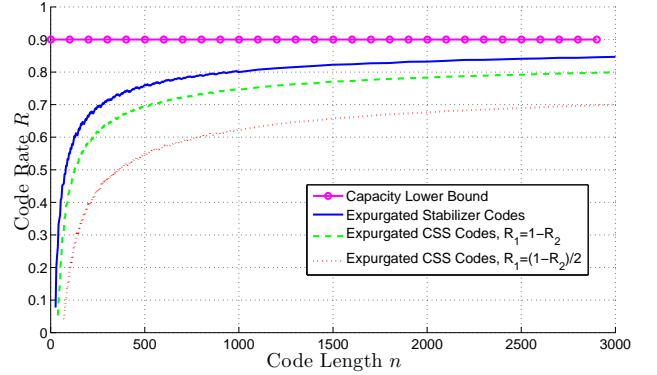}
\caption{Code Rate as the function of code length in the case of symmetric depolarizing channel with $p=0.01$ and
$1-\bar{F}\le 10^{-4}$}
\label{fig:1-Fgrowing_n}
\end{figure}

The obtained bounds allow us to estimate how the code performance
changes with the code length. In Fig.\ref{fig:1-Fgrowing_n} we
consider the case of the symmetric depolarizing channel with
$p=0.01$ and codes of
 length $n\in [30,3000]$. We choose the target fidelity as $1-F_{target}=10^{-4}$.
For a given code of length $n$ we find the largest code rate $R$
such that $1-\bar{F}\le 1-F_{target}$.  One can see that in the case
of the ensemble ${\cal Q}_{stab}$ the code rate is approaching the
capacity lower bound $\underline{C}(p)=1-2H(p)$
 ($H(p)$ is defined below in (\ref{eq:H}))
 as the code
length grows. At the same time for small code lengths, like $n=50 \mbox{ or }100$, the code rate should be significantly lower than
$\underline{C}(p)$ in order to have $1-\bar{F}\le 1-F_{target}$

Fig.\ref{fig:1-Fgrowing_n}  also shows that CSS codes have a significant data rate loss compared to stabilizer codes of the same length.
It is unclear, however, whether this loss disappears in the asymptotic regime as the code length tends to infinity. We will answer this question in the next Section.

\section{Reliability Functions and Capacity Lower Bounds}\label{sec:ErrExpon}
In this Section
we are interested in the regime when the code length $n$ tends to infinity and the code rate stays constant.

We start with studying asymptotic behavior of the average
enumerators $\hat{B}_j$ and $\hat{B}_j^\perp$ for expurgated
ensembles of stabilizer and CSS codes.

Through the rest of the paper
\begin{align}
T(x,y)&=x\log_4(3)-x\log_4(y)-(1-x)\log_4(1-y),\nonumber \\
H(x)&=T(x,x). \label{eq:H}
\end{align}

\subsection{Stabilizer and Linear Stabilizer
Codes}\label{subsec:Stab_and_Linear_Stab_Codes}

 Let ${\cal Q}_{stab}$ be the ensemble of all $[[n,k]]$ stabilizer codes.
From (\ref{Bjperp_general.stab}) and (\ref{eq:Bj_general.stab}) we
obtain
\begin{equation}
\label{eq:bw_perp_stab} \bar{b}_\omega^\perp = {1\over n}\log_4
\bar{B}^\perp_{\lfloor \omega n\rfloor}=H(\omega)-{R+1\over 2}+o(1),
\end{equation}
\begin{equation}
\label{eq:bw_stab} \bar{b}_\omega = {1\over n}\log_4
\bar{B}_{\lfloor \omega n\rfloor}=H(\omega)+{R-1\over 2}+o(1).
\end{equation}

Theorem \ref{thm:enumerators_of_Q_I} allows obtaining expurgated ensembles that give good fidelity
bounds for finite length quantum codes. In the asymptotic case it is more convenient to
form expurgated ensemble in the way
proposed in \cite{ash2000}.
For a positive $\rho>0$ we define the expurgated ensemble as
\begin{align}
&{\cal Q}_{stab,EX}\nonumber\\
=&\{Q \in {\cal Q}_{stab}: B_j(Q)-B_j^\perp (Q) \le n^{1+\rho}
(\bar{B}_j-\bar{B}_j^\perp) \mbox{ for all $j$}\}.\label{eq:Qstabex}
\end{align}
From Markov's inequality and union bound it follows that for randomly chosen $Q\in {\cal Q}_{stab}$ we have
\begin{align*}
&\Pr\left(B_j(Q)-B_j^\perp(Q)\le n^{1+\rho}
(\bar{B}_j-\bar{B}_j^\perp) \mbox{ for all $j$}\right)\\
 \ge &1-{1\over n^\rho},\rho>0.
\end{align*}
Hence for growing $n$ we have that almost any code from ${\cal Q}_{stab}$ belongs to ${\cal Q}_{stab,EX}$.

\noindent{\bf Remark} Note that the ensemble ${\cal Q}_{stab,EX}$ is different from the expurgated ensemble ${\cal Q}_{stab,ex}$
defined in Section \ref{sec:F_stab_codes}. While the polynomial factor $n^\rho$ does not affect asymptotic results (as it
will be shown later in this Section), it does not allow achieving good bounds on $1-\bar{F}$ for finite length codes. In fact, for finite
length codes bounds derived from ${\cal Q}_{stab,EX}$ are worse than similar bounds derived from the unexpurgated ensemble ${\cal Q}_{stab}$.


Denote
\begin{equation}\label{eq:GVassympt}
\delta_{GV}(R)=\lim_{n\rightarrow \infty} {d_{GV}(n,Rn)\over n}=
H^{-1}\left({1-R\over 2}\right),
\end{equation}
where $d_{GV}(n,Rn)$ is defined in Theorem \ref{thm:GV_stab}.

 From (\ref{eq:bw_perp_stab}) and (\ref{eq:bw_stab}) it follows
that
 for sufficiently large $n$ we have
$$
{1\over n}\log_4 n^{1+\rho}(\bar{B}_{\lfloor \omega n\rfloor} -
\bar{B}_{\lfloor \omega n\rfloor}^\perp)<1, \mbox{ if } \omega<
H^{-1}\left({1-R\over 2}\right).
$$
Taking into account that $B_j(Q)-B_j(Q)^\perp$ are integers, we
obtain that for any $Q\in {\cal Q}_{stab,EX}$ and sufficiently large
$n$: \begin{equation}\label{eq:BQ-BQperp,stab.EX}
 B_j(Q)-B_j(Q)^\perp=0,~1\le j\le
(\delta_{GV}(R)-\epsilon)n,  \epsilon>0.
\end{equation}
 For
$\omega\ge \delta_{GV}(R)$ we have
\begin{align}
\tilde{b}_\omega&={1\over n}\log_4 \left(B_{\lfloor \omega
n\rfloor}(Q) - {B}^\perp_{\lfloor \omega
n\rfloor}(Q)\right)\nonumber\\
&\le {1\over n} \log_4 n^{1+\rho}\left(\bar{B}_{\lfloor \omega
n\rfloor}-\bar{B}^\perp_{\lfloor \omega n\rfloor}\right) \nonumber \\
 &=\bar{b}_\omega+o(1) \nonumber \\
 &=H(\omega)+{R-1\over 2}+o(1),
 ~\omega\ge \delta_{GV}(R).\label{eq:b-b}
 \end{align}

The ensemble of ${\cal Q}_{lin.stab}$ can be analyzed in a very
similar way. Starting from the expressions (\ref{eq:stabBperp}), and
(\ref{eq:stabB}), and further defining the ensemble ${\cal
Q}_{lin.stab,EX}$ in the same way as in (\ref{eq:Qstabex}), we
obtain that equations (\ref{eq:BQ-BQperp,stab.EX}) and
(\ref{eq:b-b}) also hold for ${\cal Q}_{lin.stab,EX}$.




\subsection{CSS codes}
Now we consider the ensemble ${\cal Q}_{CSS}$  of all $(n,k_1,k_2)$ CSS codes with the average enumerators $\bar{B}_j^\perp$ and $\bar{B}_j$.

Define
$$
[n]=(2^n-1)(2^{n-1}-1)\cdot\ldots\cdot (2-1).
$$
The following identities are well known (see \cite[Ch.15.2]{macwilliams})
$$
\left[\begin{array}{c}
n\\
k
\end{array}\right]={[n]\over [k][n-k]},
$$
$$
\left[\begin{array}{c}
n\\
k_2
\end{array}\right]\left[\begin{array}{c}
k_2\\
k_1
\end{array}\right]=
\left[\begin{array}{c}
n-k_1\\
k_2-k_1
\end{array}\right]
\left[\begin{array}{c}
n\\
k_1
\end{array}\right].
$$
We use these identities in order of studying the asymptotic behavior of the ratios $c_2/c_1$ and $c_3/c_1$ for $c_1$,$c_2$, and $c_3$ defined in
Theorem \ref{thm:wt_distCSS}. We start with the first term of $c_2$ and get

\begin{align*}
&{\left[\begin{array}{c}
n-1\\
k_1-1
\end{array}\right]\left[\begin{array}{c}
n-k_1\\
k_2-k_1
\end{array}\right]\over c_1} =
{\left[\begin{array}{c}
n-1\\
k_1-1
\end{array}\right]\over
\left[\begin{array}{c}
n\\
k_1
\end{array}\right]}={[n-1][k_1]\over [n][k_1-1]}\\
=&{2^{k_1}-1\over 2^n-1}.
\end{align*}
 Similarly for the second term of $c_2$ we
have
$$
{\left[\begin{array}{c}
n-1\\
k_2
\end{array}\right]\left[\begin{array}{c}
k_2\\
k_1
\end{array}\right]\over c_1} =
{\left[\begin{array}{c}
n-1\\
k_2
\end{array}\right]\over
\left[\begin{array}{c}
n\\
k_2
\end{array}\right]}={2^{n-k_2}-1\over 2^n-1}.
$$
Hence we have
\begin{equation}\label{eq:c2overc1}
{1\over n} \log_2 \left( {c_2\over c_1}\right)=\max\{{k_1\over
n}-1,-{k_2\over n}\}+o(1).
\end{equation}
Next
 \begin{eqnarray*}
{c_3\over c_1}&=&{[n-2]\over [k_2-1][n-k_2-1]}{[k_2][n-k_2]\over [n]}\\
&&\cdot {[k_2-1]\over [k_1-1][k_2-k_1]}
{[k_1][k_2-k_1]\over [k_2]}\\
&=&{(2^{k_2}-1)(2^{n-k_2}-1)(2^{k_1}-1)\over
(2^n-1)(2^{n-1}-1)(2^{k_2}-1)}.
\end{eqnarray*}
From this we have
\begin{equation}\label{eq:c3overc1}
{1\over n}\log_2 \left({c_3\over c_1} \right)= {k_1\over
n}-1-{k_2\over n}+o(1).
\end{equation}

Now, denoting by
$$R_1=\log_2 {|C_1|\over n}\mbox{ and }
R_2=\log_2 {|C_2|\over n}$$
the codes rates of $C_1$ and $C_2$
respectively, we obtain the following Theorem.

\begin{theorem}\label{thm:br}
\begin{align*}
&\bar{b}_\xi^\perp = {1\over n}\log_4 \bar{B}^\perp_{\lfloor \xi n\rfloor}\\
=&H(\xi)-\xi\log_4(3)\\
+&\max\left\{ {R_1-1\over 2},-{R_2\over
2},\xi\log_4(3)+{R_1-1-R_2\over 2}\right\} +o(1),
\end{align*}
\begin{align*}
&\bar{b}_\xi = {1\over n}\log_4 \bar{B}_{\lfloor \xi n\rfloor}\\
=&H(\xi)-\xi\log_4(3)\\
+&\max\left\{{R_2-1\over 2},-{R_1\over 2},\xi\log_4(3)+{
R_2-1-R_1\over 2}\right\}+o(1).
\end{align*}
\end{theorem}

For a positive $\rho>0$, we define the expurgated ensemble of CSS codes as
 \begin{align}
{\cal Q}_{CSS,EX} =&\{Q\in {\cal Q}_{CSS}: \nonumber\\
& B_j(Q)-B_j^\perp(Q)  \le n^{1+\rho}
(\bar{B}_j-\bar{B}_j^\perp)\}.\label{eq:QCSSex}
\end{align}

\noindent{\bf Remark} Again we note that the ensemble ${\cal Q}_{CSS,EX}$ is different from ${\cal Q}_{CSS,ex}$
defined in Section \ref{sec:F_CSS_codes}. The ensemble ${\cal Q}_{CSS,EX}$ is convenient for asymptotic analysis,
but gives bad results for finite length CSS codes.

Theorem \ref{thm:br} allows us to find the asymptotic expression for
$d_{GV,CSS}(n,k_1,k_2)$ from Theorem \ref{thm:GV_CSS}. Let
$$
\delta_{GV,CSS}(R_1,R_2)=\lim_{n\rightarrow \infty}
{d_{GV,CSS}(n,R_1n,R_2n)\over n}.
$$
Taking into account that $B(Q)-B(Q)^\perp$ are integers and that
$\bar{b}_\xi>\bar{b}_\xi^\perp,\xi\in (0,1)$, we conclude that
$\delta(R_1,R_2)$ is the root of the equation $\bar{b}_\xi=0$. After
simple calculations we obtain
\begin{equation}\label{eq:GVassympt_CSS}
\delta_{GV,CSS}(R_1,R_2)=H_2^{-1}(\min(R_1,1-R_2)),
\end{equation}
where
$$
H_2(x)=-x\log_2(x)-(1-x)\log_2(1-x)
$$
is the binary entropy. Note that if $R_1=1-R_2$ then $R_1=(1-R)/2$
we get the usual Gilbert-Varshamov bound for CSS codes
\cite[Ch.10.4.2]{nielsen}:
$$
\delta_{GV,CSS}(R)=H_2^{-1}\left({1-R\over 2}\right).
$$
Summarizing this, we obtain the following result.
\begin{theorem}
Let $Q\in {\cal Q}_{CSS,EX}$. Then
\begin{enumerate}
\item for any $\epsilon>0$ and sufficiently large $n$ we have
 \begin{equation} \label{eq:B-B_CSS}
B_j(Q)-B_j^\perp(Q)=0,~1\le j \le
(\delta_{CSS,GV}(R_1,R_2)-\epsilon)n,
\end{equation}
\item for $\omega>\delta_{CSS,GV}(R_1,R_2)$ we have
\begin{equation}\label{eq:bcss-bcss}
\tilde{b}_\omega={1\over n}\log_4 ({B}_{\lfloor \omega n\rfloor}(Q)
- {B}^\perp_{\lfloor \omega n\rfloor}(Q))\le\bar{b}_\omega+o(1),
\end{equation}
where $\bar{b}_\omega$ is defined in Theorem \ref{thm:br}.
\end{enumerate}
\end{theorem}

\begin{figure}[htb]
\centering
\includegraphics[scale=0.38]{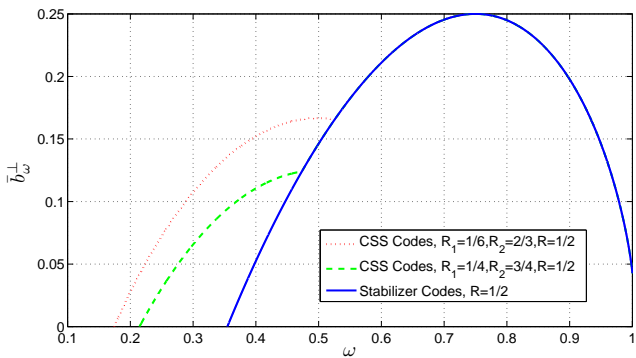}
\caption{Average Quantum Enumerators $\bar{b}_\omega^\perp$ of
Stabilizer and CSS codes (only positive part of
$\bar{b}_\omega^\perp$ is shown) } \label{fig:bdual}
\end{figure}

\begin{figure}[htb]
\centering
\includegraphics[scale=0.38]{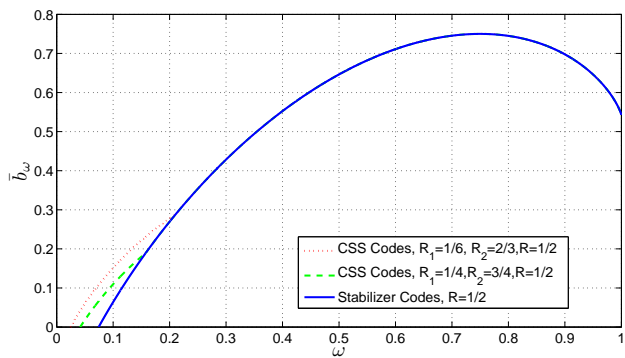}
\caption{Average Quantum Enumerators $\bar{b}_\omega$ of Stabilizer
and CSS codes (only positive part of $\bar{b}_\omega$  is shown)}
\label{fig:b}
\end{figure}

Examples of average quantum enumerators for stabilizer and CSS codes
are shown in Fig.\ref{fig:bdual} and Fig.\ref{fig:b}. The figures
show that at certain range of $\omega$ quantum enumerators of CSS
codes are exponentially larger than their stabilizer counterparts.
We see in the next subsection that this leads to an exponential loss
of performance of CSS codes.

\subsection{Reliability Functions}
Let ${\cal Q}$ be an ensemble of quantum codes of rate $R$ and length $n$. For the quantum depolarizing channel with the channel probability $p$ define
$$
E(n,R,p)=\sup_{Q\in {\cal Q}} -{1\over n} \log_4 (1-F(Q))
$$
The {\em reliability function}, which is also called the error
exponent, of ${\cal Q}$ is defined as
$$
E(R,p)=\lim \inf_{n\rightarrow \infty} E(n,R,p).
$$
When $E(R,p)$ is positive it shows how fast $1-F$ approaches zero as
$n\rightarrow \infty$. Thus for given $R$ and $p$ we want that
$E(R,p)$ be as large as possible.

In what follows we will compute lower bounds on $E(R,p)$ for the ensembles of stabilizer and CSS codes. In order of doing this we
analyze the exponent of (\ref{eq:poltbound_ensembl}) with the quantum enumerators of stabilizer and CSS codes.

We start with computing the exponent of $G(m,w)$ defined in (\ref{eq:G(m,w)}). Let
$$
\mu={m\over n},\omega={w\over n},\eta={h\over n}, \mbox{ and } \tau={t\over n}.
$$
For the $h$-th term of the second sum of $G(m,w)$ we have
 \begin{eqnarray}
&&{1\over n}\log_{4} {w-t\choose h}{n-w\choose m-t-h}3^{m-t-h}\nonumber \\
&=& (\omega-\tau)H({\eta\over \omega-\tau})-\eta\log_4(3)\nonumber\\
&+&(1-\omega)H({\mu-\tau-\eta\over 1-\omega})+o(1).\label{eq:eta1}
\end{eqnarray}
The roots of the derivative of this expression with respect to
$\eta$ are
$$
\eta_1=a+{1\over 4}\sqrt{b},~\eta_2=a-{1\over 4}\sqrt{b}$$
where
 $$
a={\omega\over 2}-{\tau\over 4}+{\mu\over 2}-{3\over 4}
$$
and
\begin{align*}
b&=4\omega^2-12\omega\tau+16\omega\mu-12\omega\\
&+9\tau^2-12\tau \mu+6\tau+4\mu^2-12\mu+9.
\end{align*}
It is not difficult to show that the maximum of (\ref{eq:eta1}) is
achieved at $\eta_1$ (in fact $\eta_2$ is always negative when
$\mu$, $\omega$, and $\tau$ belong to their summation ranges).
Further one can show that $\eta_1<(\omega-\tau)/2$ when $\mu$,
$\omega$, and $\tau$ belong to their summation ranges. Thus the
maximum of (\ref{eq:eta1}) is achieved at
$$
\eta^*=(\omega-\tau)/2.
$$
Using $\eta^*$ we compute the exponent of the $t$-th term of the first sum of $G(m,w)$
and obtain
 \begin{eqnarray*}
&&{1\over n}\log_4 \left({w\choose t}2^t {w-t\choose {w-t\over 2}}{n-w\choose {2m-t-w\over 2}}3^{{2m-t-w\over 2}}\right)\\
&=& \omega H\left({\tau\over
\omega}\right)-\tau\log_4(3)+{\omega\over
2}\\
&&+(1-\omega)H\left({2\mu-\tau-\omega\over 2(1-\omega)}\right).
\end{eqnarray*}
The derivative of this expression in $\tau$ has three roots. Two of
them take either negative or complex values when $\mu$ and $\omega$
belong to their summation ranges. The root that takes nonnegative
values is
$$
\tau^*={(a+36\sqrt{b})^{1/3}\over 12}-{12c\over ((a+36\sqrt{b})^{1/3})}+{1\over 3}\omega+{2\over 3}\mu-{1\over 2},
$$
where
 \begin{eqnarray*}
a&=&672\mu\omega^2+192\omega\mu^2-720\omega\mu-80\omega^3-396\omega^2\\
&&+432\omega+512\mu^3-1152\mu^2+864\mu-216;\\
b& = &24\mu\omega^3+528\mu^2\omega^2-147\omega^4-144\mu\omega^2+72\omega^3-12\omega^6\\
&&+136\omega^5+192\omega^4\mu^2-112\omega^5\mu-96\omega^3\mu^2-48\omega^4\mu\\
&&-640\omega^2\mu^3+256\omega^2\mu^4;\\
c &=& -{1\over 9}\omega\mu-{7\over 36}\omega^2+{1\over
3}\omega-{4\over 9}\mu^2+{2\over 3}\mu-{1\over 4}.
\end{eqnarray*}
 We omit large analytical expressions for other two roots.

Denote
\begin{align*}
f(\mu,\omega)=&\omega H\left({\tau\over \omega}\right)-\tau^*\log_4(3)\\
&+{\omega\over 2}+(1-\omega)H\left({2\mu-\tau^*-\omega\over
2(1-\omega)}\right).
\end{align*}

\noindent{\bf Remark} As we noted in the end of section
\ref{subsec:Stab_and_Linear_Stab_Codes}, the ensembles ${\cal
Q}_{stab,EX}$ and ${\cal Q}_{lin.stab,EX}$ have the same
$\tilde{b}_\omega$. From this it follows that these ensembles have
the same reliability function. Therefore all results presented below
for stabilizer codes are also valid for linear stabilizer codes.

For the ensembles ${\cal Q}_{stab,EX}$ and ${\cal Q}_{CSS,EX}$
defined in (\ref{eq:Qstabex}) and (\ref{eq:QCSSex}) respectively and
for $\bar{N}(m)$ defined in (\ref{eq:N(m)}) we have
$$
{1\over n}\log_4 N(\lfloor \mu n\rfloor )=\max_{\delta<\omega<2\mu}
\{ \tilde{b}_\omega+f(\mu,\omega)\},
$$
and further, according to (\ref{eq:poltbound_ensembl}),
 \begin{align*}
E(R,p) &\ge\underline{E}(R,p)=\max_{\delta/2<\mu<1} \left\{\mu \log_4(p/3)\right.\\
&+(1-\mu)\log_4(1-p)\nonumber \\
&+\left.\min\{H(\mu),\max_{\delta<\omega<2\mu}
\{\tilde{b}_\omega+f(\mu,\omega)\}\right\}
\end{align*}
where $\delta$ is either $\delta_{GV}(R)$ or
$\delta_{GV,CSS}(R_1,R_2)$ and $\tilde{b}_\omega$ is defined either
in (\ref{eq:b-b}) or (\ref{eq:bcss-bcss}) respectively.

Numerical computation of $\underline{E}(R,p)$, for both ${\cal
Q}_{stab,EX}$ and ${\cal Q}_{CSS,EX}$, is easy. The function inside
of $\max_{\delta <\omega <2 \mu }\{ \cdot \}$ and the function
inside of $\max_{\delta/2< \mu <1} \{ \cdot \}$ were found to be
concave by inspection, for parameter values tested. In the case of
${\cal Q}_{stab,EX}$ numerical computations show that
$\underline{E}(R,p)$ coincides with Gallager's exponent
(reliability function) of random classical quaternary code of rate
$(R+1)/2$. It also coincides with the reliability function found in
\cite{hamada2001},\cite{hamada2002},\cite{barg}:

 \begin{align*}
&E_{stab}(R,p)\ge \underline{E}(R,p)\\
=&\left\{\begin{array}{ll} -\delta_{GV}\left({R+1\over
2}\right)\log_4\left(\sqrt{{4\over
3}p(1-p)}\right. & \\
\left.+{2\over 3}p\right),& 0\le R\le R_{min}\\
1-\log_4(1+2p+\sqrt{12p(1-p)}) & \\
-{R+1\over 2}, & R_{min}\le R\le R_{cr}\\
T(\delta_{GV}\left({R+1\over 2}\right),p)-1+{R+1\over 2}, & R_{cr}\le R\le \underline{C}(p) \\
\end{array}\right.
\end{align*}
where
\begin{align*}
\underline{C}(p)&={1-2H(p)},\\
R_{cr}&=\max\left\{0,1-2H\left({\sqrt{3p}\over
\sqrt{3p}+\sqrt{1-p}}\right)\right\},\\
R_{min}&=\max\left\{0,1-2H\left({3\alpha\over
1+3\alpha}\right)\right\},
\end{align*}
and
$$
 \alpha=\sqrt{{4\over 3}p(1-p)}+{2\over
3}p.
$$

\begin{figure}[htb]
%
\hspace{-0.4cm}\includegraphics[scale=0.36]{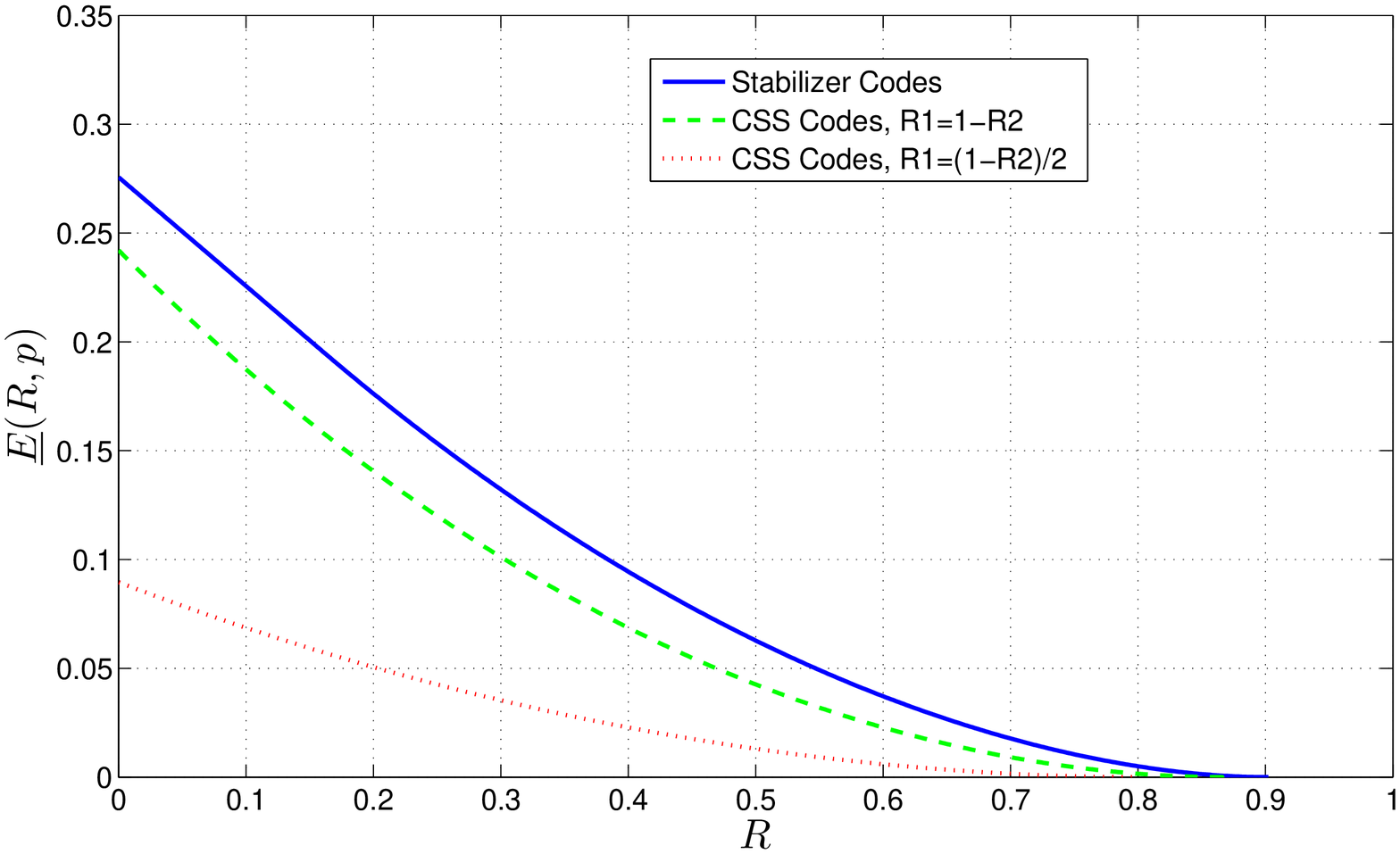}
\caption{Error Exponents for Random Stabilizer and CSS Codes in
symmetric depolarizing channel with $p=0.01$ } \label{fig:err_exp}
\end{figure}

The reliability function for the depolarizing channel with
 $p=0.01$ are shown in Fig.\ref{fig:err_exp}. One can see that stabilizer codes slightly outperformed balanced CSS codes ($R_1=1-R_2$),
 and that balanced CSS codes significantly outperform unbalanced CSS codes with $R_1=(1-R_2)/2$.

\begin{figure}[htb]
\centering
\includegraphics[scale=0.38]{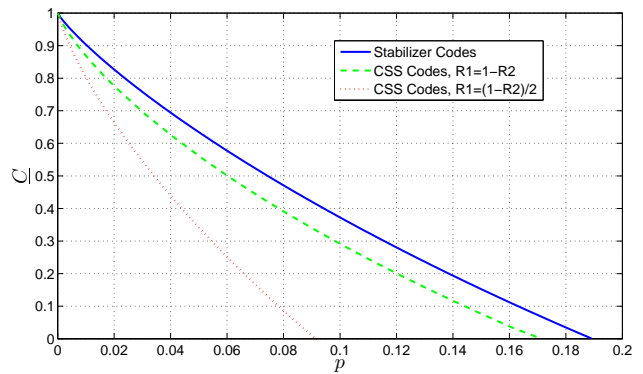}
\caption{Capacity Lower Bounds for Stabilizer and CSS Codes}
\label{fig:lower_b_capacity}
\end{figure}

We can use the function $\underline{E}(R,p)$ for obtaining a lower
bound $\underline{C}(p)$ on the capacity of an ensemble ${\cal Q}$
of quantum codes. For an ensemble ${\cal Q}$ of quantum codes and
for given channel error $p$  we define $\underline{C}(p)$ by
$$
\underline{C}_{\cal Q}(p)=\max\{R: \underline{E}(R,p)>0\}.
$$
The capacity lower bounds for stabilizer, balanced and unbalanced
CSS codes are shown in Fig.\ref{fig:lower_b_capacity}. We see again
that stabilizer codes slightly outperform balanced CSS codes, and
that CSS codes with $R_1=(1-R_2)/2$ are significantly weaker than
stabilizer and balanced CSS codes.

\section{Appendix}

\IEEEproof (Lemma \ref{lem:1})  Any $C_2^\perp$ uniquely defines the code $C_2$. So it is enough to count in how many ways code $C_2^\perp$ can be constructed.
Any vector from $C_2^\perp$ should be orthogonal to $\bv$. There are $(2^{n-1}-1)$ nonzero vectors that are orthogonal to $\bv$.
We use $\bw$ as the first basis vector of $C_2^\perp$. Other
$n-k_2-1$ basis vectors can be chosen in
$$
(2^{n-1}-2)(2^{n-1}-2^2)\ldots (2^{n-1}-2^{n-k_2-1})
$$
ways. In any such code its basis vectors can be chosen in
$$
(2^{n-k_2}-2)(2^{n-k_2}-2^2)\ldots (2^{n-k_2}-2^{n-k_2-1})
$$
ways. The ratio of the above expressions is equal to
$$
\left[\begin{array}{cc}
n-2\\
n-k_2-1
\end{array}\right]=\left[\begin{array}{cc}
n-2\\
k_2-1
\end{array}\right]
$$
and it defines the number of ways in which code $C_2^\perp$ and code
$C_2$ can be constructed. \qed

\IEEEproof (Lemma \ref{lem:C2s that contain given C1}) Any $k_1$ basis vectors of the $[n,k_1]$ code can be chosen as the fist $k_1$ basis vectors of an $[n,k_2]$ code.
Other $k_2-k_1$ vectors can be chosen in
$$
(2^{n}-2^{k_1})(2^{n}-2^{k_1+1})\ldots (2^{n}-2^{k_2-1})
$$
ways. In any such $[n,k_2]$ code its $(k_2-k_1)$ basis vectors can be chosen in
$$
(2^{k_2}-2^{k_1})(2^{k_2}-2^{k_1+1})\ldots (2^{k_2}-2^{k_2-1})
$$
ways. The ratio of the above expressions completes the proof. \qed

\IEEEproof (Lemma \ref{lem:C1s codes in a given C2})  The number of nonzero vectors in an $[n,k_2]$ code is $2^{k_2}-1$. Hence a basis of an $[n,k_1]$ code can be chosen in
$$
(2^{k_2}-1)(2^{k_2}-2)\ldots (2^{k_2}-2^{k_1-1})
$$
ways. In any such $[n,k_1]$ code we can choose a basis in
$$
(2^{k_1}-1)(2^{k_1}-2)\ldots (2^{k_1}-2^{k_1-1})
$$
ways. The number of distinct $[n,k_1]$ codes in a given $[n,k_2]$ code is equal to the ratio of the above expressions. \qed

\IEEEproof (Lemma \ref{lem:C1s_with_vector_a_in_given_C2}). The
proof is similar to the proof of Lemma \ref{lem:C1s codes in a given
C2}.

\IEEEproof (Lemma \ref{lem:w in C2perp}) Nonzero vectors from $C_2$
must be orthogonal to $\bw$. There are $2^{n-1}-1$ such vectors.
Hence $k_2$ basis vectors of $C_2$ can be chosen in
$$
(2^{n-1}-1)(2^{n-1}-2)\cdot\ldots\cdot(2^{n-1}-2^{k_2-1})
$$
ways. In a given $[n,k_2]$ code a basis can be chosen in
$$
(2^{k_2}-1)(2^{k_2}-2)\cdot\ldots\cdot(2^{k_2}-2^{k_2-1})
$$
ways. The ratio of the above expressions competes the proof.
\qed

\section*{acknowledgement}
This research was supported by the Intelligence Advanced Research
Projects Activity (IARPA) via Department of Interior National
Business Center contract number D11PC20165. The U.S. Government is
authorized to reproduce and distribute reprints for Governmental
purposes notwithstanding any copyright annotation thereon. The views
and conclusions contained herein are those of the authors and should
not be interpreted as necessarily representing the official policies
or endorsements, either expressed or implied, of IARPA, DoI/NBC, or
the U.S. Government.

\end{document}